\documentclass[conference]{IEEEtran}
\usepackage{cite}
\usepackage{amsmath,amssymb,amsfonts}
\usepackage{algorithmic}
\usepackage{graphicx}
\usepackage{textcomp}
\usepackage{xcolor}
\usepackage{dblfloatfix}
\usepackage[colorinlistoftodos]{todonotes}
\def\BibTeX{{\rm B\kern-.05em{\sc i\kern-.025em b}\kern-.08em
    T\kern-.1667em\lower.7ex\hbox{E}\kern-.125emX}}
\begin{document}
\bstctlcite{IEEEexample:BSTcontrol}

\title{Rapid Method for Generation Prioritization during System Restoration with Renewable Resources}

\author{\IEEEauthorblockN{Adam Mate, Eduardo Cotilla-Sanchez}
\IEEEauthorblockA{\textit{School of Electrical Engineering \& Computer Science} \\
\textit{Oregon State University, Corvallis, OR 97331 USA}\\
matea@oregonstate.edu, ecs@oregonstate.edu}}

\maketitle

\begin{abstract}
Quick and reliable power system restoration is critically important after natural disasters or other sudden threats, such as cyber-attacks.
Leveraging renewable resources in system restoration shortens recovery times, resulting in prevented life-loss and avoided economic-loss, and improves the resilience of the entire grid. However, it is not a common practice today; the inherent variability of these resources represents a challenge for a streamlined restoration process.
This paper presents a prioritized method -— starting with renewable generator units then lowering priority to conventional units -— to plan the operational schedule of a power system during the restoration process. The goal is to achieve a well balanced system in the presence of significant renewable penetration.
Validation and benchmarking experiments were performed on a customized version of the RTS-GMLC test system using six months out of year-long data, tested through hourly simulations. After evaluating the performance and computational costs, this method proved faster than common approaches:~a MILP Unit Commitment algorithm, widely used today, and an ``enable-and-try'' algorithm. In summary, herein a more convenient method is provided to be utilized during time-sensitive restoration, as an online operation-planning aid.
\end{abstract}


\section{Introduction}

Natural disasters (e.g. hurricanes, earthquakes, floods) and other extreme weather conditions (e.g blizzards, heat waves) are becoming more common, posing an increasing threat to our power systems.
The U.S.~Pacific Northwest (PNW) in particular faces a complex and devastating disaster scenario:~the imminent Cascadia Subduction Zone (CSZ) megathrust earthquake, yielding the creation of a powerful tsunami, hundreds of aftershocks and increased volcanic activity in the region \cite{intro_1} --\cite{intro_2}.
To effectively cope with such challenges, the resiliency of the power system must be improved and the speed of power system restoration accelerated.
Several types of renewable generators are able to withstand the above threats better than traditional ones. Wind turbines and solar panels have proven many times their ability to quickly respond to and recover from extreme events \cite{intro_3}. Therefore, renewable resources should be leveraged during restoration as they are convenient to use to shorten recovery times \cite{intro_4} --\cite{intro_5}.

A key step in power system restoration is determining the operational schedule of restored units.
Classic unit commitment (UC) algorithms optimize for operational costs (i.e., supply system loads with the lowest total cost), however, they can take an extended amount of time to find the optimal schedule depending on the integrated modeling details and the selected solution approach (for details see Section \ref{M-Two}).

After catastrophes, every minute counts in system restoration. Any delay can have tragic consequences.
In this paper a prioritization method is proposed for power systems with significant renewable penetration; greater than 20\% average. The method prioritizes generator units, and determines which ones should be dispatched. Renewable units that remain available after the event are enabled by default, and the goal is to decide within seconds which conventional units (and with what generation setpoints) should be enabled alongside them.
As data is received about the ongoing restoration process, the method can reconsider its earlier made decisions close to real time and adjust to always provide the best schedule.

\section{Model Description} \label{MD}


To evaluate the proposed prioritization method, the \textit{Reliability Test System of the Grid Modernization Lab Consortium} (RTS-GMLC) test system \cite{rts-gmlc_1} was used.
RTS-GMLC is an updated version of the IEEE RTS-96 test system \cite{rts-gmlc_2}, with modernizing changes \cite{rts-gmlc_3}:
\begin{itemize}
\item Created relative node locations based on line distances (arbitrary geographic region in the SW United States).
\item Fixed data errors, improved transmission system, updated bus loads and regional load profiles, modernized generation fleet (new unit types, new conventional generators, and new renewable generation profiles).
\item Hourly and 5-minute operations data for a year -- from Jan.~1st, 2020 to Dec.~31st, 2020.
\end{itemize}

The default RTS-GMLC case represents a peak load flow state, with disabled wind and solar generations. It consists of 73 buses, 158 generator units (including 72 conventional and 82 renewable units), 120 AC transmission lines, and 51 loads. Case-values are based on the original RTS-96 system.
Forecasted hourly data is available for the active power generation of solar, wind and hydro units, and for the active loads of the system. Hourly data is not provided for the reactive power generations and loads, conventional units, synchronous condensers (abbrv.: sync-conds), and the storage unit.


The below used term \textit{time\_period} refers to a single hour period of the year period: the RTS-GMLC data-set consists of 8,784 hour-sized \textit{time\_period}s.

\vspace{2mm}
The following subsections present and discuss in detail all implemented customization of the RTS-GMLC test system.

\subsection{Load Data} \label{LD}


Hourly real power demand data is provided for each area and in every \textit{time\_period} separately.
To get the new $P_d$ load value (\textit{newMW}) of a specific bus in an area:
\begin{equation}
\textnormal{newMW} = \textnormal{oldMW} \cdot \textnormal{MW\_rescaling}
\end{equation}
where \textit{oldMW} is the default RTS-GMLC active load value of the bus, and \textit{MW rescaling} ratio-value is determined as:

\begin{equation}
\textnormal{MW\_rescaling} = \frac{\textnormal{time\_period\_load}}{\textnormal{total\_demand}}
\end{equation}
where \textit{time\_period\_load} is the provided active load timeseries value of the area, and \textit{total\_demand} is the calculated total real power demand of the entire area in the \textit{time\_period}.


\vspace{2mm}
Hourly reactive power demand data is not provided.
The default $Q_d$ load values of buses were kept unchanged from the RTS-96 values. Using a fixed, peak load flow state value during the simulations, however, is not an accurate characterization of the reactive load profile that varies throughout a day and the year. Thus, new rescaling ratio-values were introduced to improve the load profile.
To get the new $Q_d$ load value (\textit{newMVar}) of a specific bus in an area:
\begin{equation}
\textnormal{newMVar} = \textnormal{oldMVar} \cdot \textnormal{MVar\_rescaling}
\end{equation}
where \textit{oldMVar} is the default RTS-GMLC reactive load value of the bus, and \textit{MVar\_rescaling} ratio-value is determined as:

\begin{equation}
\textnormal{MVar\_rescaling}= \frac{\textnormal{time\_period\_load}}{\textnormal{max\_demand}} 
\end{equation}
where \textit{time\_period\_load} is the provided active load timeseries value of the area, and \textit{max\_demand} is the determined maximum \textit{time\_period\_load} value of the area throughout the entire year.
More specifically, to create the new \textit{MVar\_rescaling} ratio-value of an area in a certain \textit{time\_period}, the provided active load timeseries values were used:
\textit{1)} the yearly maximum timeseries value of the area is determined and set to 1, and 
\textit{2)} the values of other \textit{time\_period}s are the calculated ratios compared to the area-maximum.


\begin{figure}[htbp]
\centerline{\includegraphics[scale=0.45]{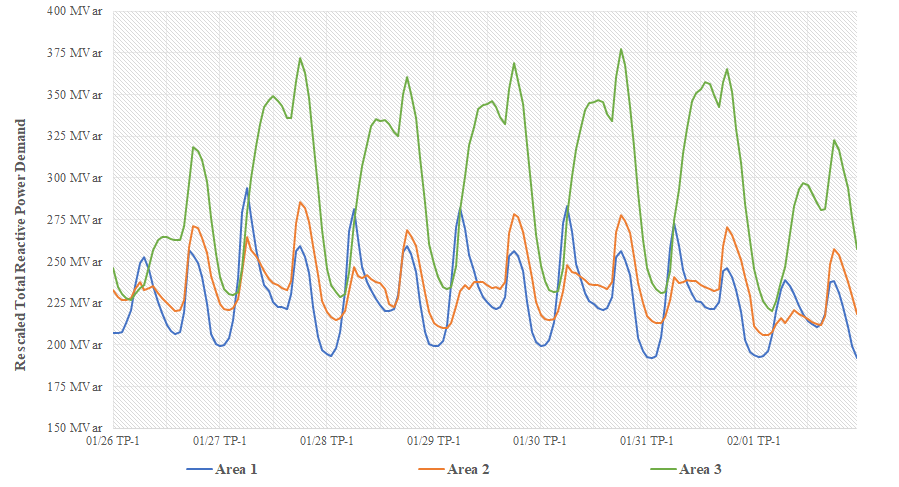}}
\caption{Improved reactive load profile of the RTS-GMLC system.}
\label{MVar_re-scaling}
\end{figure}
\vspace{5mm}

Fig.~\ref{MVar_re-scaling} presents the improved reactive load profile of RTS-GMLC between January 26 and February 2 (i.e. 168 \textit{time\_period}s).
The graph illustrates how the total power demand of each area changes throughout the days and the week, instead of staying at constant 580 [MVar] values. This is a more realistic profile, a better fit for the hourly simulations.

\subsection{Energy Portfolio} \label{EP}


\begin{figure*}[!b]
\centerline{\includegraphics[scale=0.275]{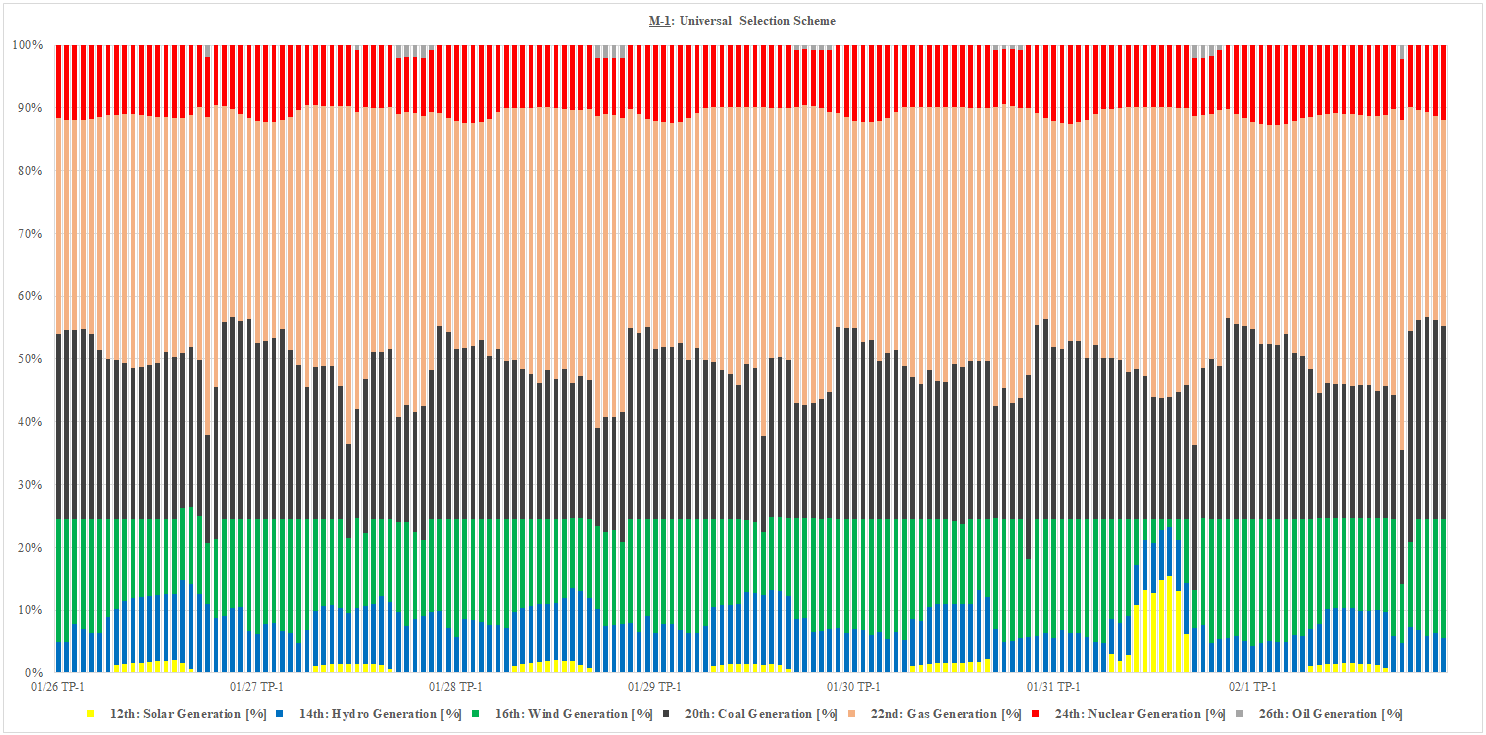}}
\vspace{1mm}
\centerline{\includegraphics[scale=0.275]{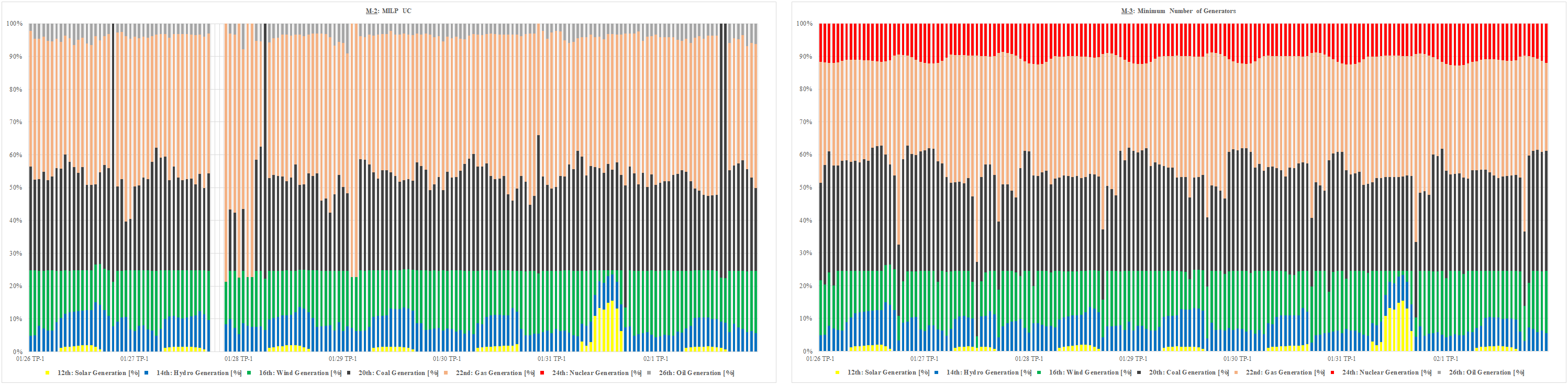}}
\caption{Modified portfolio of the RTS-GMLC system, using different prioritization approaches.}
\label{New_Energy_Portfolios}
\end{figure*}



In the RTS-GMLC system, the forecasted renewable generation is substantial throughout the year, and could supply most loads by itself in numerous \textit{time\_period}s.
Enabling every generators (fixed-value conventional and hourly-changing renewable units) simultaneously would lead to an unbalanced power system where the generation greatly exceeds the demand. For this reason, the available units must be coordinated; key generators need to be selected to operate based on the system state and operational goals.


The renewable generation profile of RTS-GMLC is based on the Southwest U.S., a region filled with solar and wind resources \cite{rts-gmlc_1}.
Today's energy portfolio of Oregon and the PNW, however, differs from this: about a half and a tenth of the generated power comes from hydro and wind resources, respectively \cite{energy_1} --\cite{energy_2}.
The desired goal was to change the renewable portfolio of RTS-GMLC to resemble the PNW's portfolio in every \textit{time\_period}.
In this research, based on historical data and anticipated generation-changes, the renewable portfolio of the PNW (i.e. \textit{goal portfolio}) in 2020 is predicted to be: Solar 0.5\%, Hydro 46.75\%, Wind 10.5\%, and Other Renewable 2.25\%. The minimum renewable generation requirement in 2020, based on legal mandates \cite{energy_3} --\cite{energy_4}, is assumed to be 20\%.

\vspace{2mm}
In achieving significant renewable penetration (that characterizes the PNW), the forecasted available potentials of every renewable generators were modified in each \textit{time\_period}:
\begin{itemize}
\item Renewable generators are the highest priority; all units are enabled that are forecasted to have generation.
Concentrated Solar Power (CSP) units are not used in the PNW, so were disabled.
\item The 2.25\% ``Other Renewable'' is proportionally distributed between the ``Wind'' and ``Solar'' categories of \textit{goal portfolio}. Distribution is based on the current generation's share of the total renewable generation.
\item After the total active power demand (\textit{total\_load}) is determined, the allowed generation (based on forecasted data) of renewable units is changed:
\begin{itemize}
\item units of a specific resource-type collectively generate $P_{total-type}$ active power
\item individual units of that type can only generate as much power ($P_{gmax}$) as their $P_{total-type}$'s share of the \textit{total\_load} is less (or equal) to the corresponding percentage in the \textit{goal portfolio}
\item if the share is less than the \textit{goal portfolio} percentage, their $P_{gmax}$ limits are set to be their forecasted generations in that \textit{time\_period}; otherwise $P_{gmax}$ limits are rescaled to achieve the \textit{goal portfolio}
\end{itemize}
\item In case the minimum renewable generation requirement is not fulfilled at this point, and potential is remained in the forecasted maximum generation, the allowed generation ($P_{gmax}$) of appropriate units are increased with the remainder generations to achieve requirement fulfillment.
\item $P_{gmin}$ limits of renewable units are kept unchanged.
Since hourly data is not provided for reactive power generations, the default RTS-GMLC values were kept unchanged throughout the year.
Solar and wind units are not able to participate in reactive power generation.
\end{itemize}

\subsection{Synchronous Condensers and Storage Units} \label{SC-SU}


Reactive power is not able to travel far, thus it must be generated where it is used.
The renewable generators of RTS-GMLC greatly contribute towards the active power supply of loads, but are not contributing towards the reactive power generation (except for hydro units).
With the introduced significant renewable share and preferred use of renewable units, this leads to cases with insufficient reactive power generation, and with substantial generation imbalance across the power system.
For this reason, the sync-conds of the default RTS-GMLC case must be reassessed and modified.

Hourly data is not provided for the sync-conds.
The default reactive power generation limits of the three existing sync-conds (one in each area: Bus114, Bus214 and Bus314) were updated to better fit the changed energy portfolio of the system: $Q_{gmin}$ minimum limit was set to -50 [MVar], and $Q_{gmax}$ maximum limit was set to 100 [MVar].
Furthermore, to compensate for the missing reactive power generation and slightly reduce the system-level imbalance, new sync-conds were added with realistic generation-limits.

In each \textit{time\_period}, every bus that has a connected renewable unit receives a new added sync-cond. The generation limits of these additional sync-conds are based on the total power generation of the buses renewable unit(s).
If the total generated active power of the unit(s) is greater than 250 [MW], then the limits of the added sync-cond are set as: $Q_{gmin}$ is -50 [MVar], and $Q_{gmax}$ is 100 [MVar].
If the total generated power is greater than 100 [MW], then the limits set as: $Q_{gmin}$ is -25 [MVar], and $Q_{gmax}$ is 25 [MVar].
Otherwise, $Q_{gmin}$ is -5 [MVar], and $Q_{gmax}$ is 10 [MVar].


\vspace{2mm}
The storage unit of the system was disabled.




\subsection{Observations} \label{Observations}

Fig.~\ref{New_Energy_Portfolios} presents the modified energy portfolio of RTS-GMLC between January 26 and February 2, using different generation prioritization approaches.
The Universal Selection Scheme (abbrv.: USS -- Section \ref{M-One}; top graph) is the proposed new method, while the MILP Unit Commitment (abbrv.: MILP UC -- Section \ref{M-Two}; bottom left graph) and the Minimum Number of Generators (abbrv.: MNG -- Section \ref{M-Three}; bottom right graph) algorithms were implemented for result comparison. The graphs illustrate the similarities and differences between the determined operational schedules.

Performed experiments (Section \ref{Results}) verify that the changed portfolio of RTS-GMLC has significant renewable penetration.
Average 20-25\% of the total system load is supplied by renewable sources throughout the \textit{time\_period}s, so the minimum renewable generation requirement is fulfilled.
Also, the modified portfolio of the system broadly resembles the PNW's predicted energy portfolio.
On the other hand, there are notable differences between the used prioritization approaches; these are discussed in detail in Section \ref{Results-Normal}.

\newpage

\section{Generation Prioritization Method} \label{GPM}

Beside the significant renewable generation in the modified RTS-GMLC (Section \ref{EP}), the remainder of the total system load is supplied by nonrenewable sources. Thus, the conventional generators must be prioritized and key units selected to generate. The following subsections present a new prioritization method for this process.
\vspace{2mm}

\subsection{GPWD Factor} \label{GPWD}

To characterize the importance of each generator, the \textit{Generator Participation Weight Determination} (GPWD) factor was introduced.
This new index is comprised of easily obtainable values, and is used to rank generators in each \textit{time\_period}, creating a list that distinguishes between significant and less significant units.
The formed list has a vague resemblance to priority lists presented in \cite{pl_1} or \cite{pl_2}, but is more relevant to be used during time-sensitive restoration than those.

\vspace{2mm}
GPWD is calculated as follows:
\begin{equation}
\textnormal{GPWD} = \textnormal{PS} + \textnormal{APF-P} + \textnormal{APF-Q} + \textnormal{MP} - 
\frac{\textnormal{$P_{gmin}$}}{\textnormal{$Q_{gmax}$}}
\end{equation}

\textit{PS}: Prior State
\begin{itemize}
\item the Status of a unit in the prior \textit{time\_period}; enabled unit receives a value of 1, disabled unit receives 0
\item turning generators ON and OFF frequently is not beneficial or realistic, so previously enabled units have higher rank in the present \textit{time\_period}
\end{itemize}

\vspace{2mm}
\textit{APF-P} and \textit{APF-Q}
\begin{itemize}
\item Area Participation Factors based on $P_g$ active power, and $Q_g$ reactive power generations; values of enabled units add up to 1 in each area for both cases
\item To obtain values: After setting $P_{gmin}$ generation limits of all conventional generators to 0 [MW], an Optimal Power Flow (OPF) \cite{opf_1} simulation is performed. Then, using the OPF results in every area separately: \textit{1)} calculate the total \textit{Pg} (or absolute valued $Q_g$) generation of conventional units; \textit{2)} determine each individual unit's share of the total generation.
\item higher \textit{APF} value means greater contribution in the area, resulting in a more important unit
\end{itemize}

\vspace{2mm}
\textit{MP}: Maximum Power
\begin{itemize}
\item maximum generatable power of a certain unit compared to the largest generator of the power system; each unit receives a value between 0 and 1
\begin{itemize}
\item if the relative $P_{gmax}$ (or $Q_{gmax}$) size of a unit is greater than 95\%, the unit receives a value of 0.5; if $P_{gmax}$ (or $Q_{gmax}$) is between 95\% and 80\%, the unit receives 0.25; otherwise the unit receives 0
\item \textit{MP} is the sum of the two values resulting from the relative $P_{gmax}$ and $Q_{gmax}$ sizes; only the largest units in the power system receive \textit{MP} values
\end{itemize}
\item \textit{MP} keeps the largest unit(s) of the system active most of the time, as they greatly contribute towards the missing load supply, and are harder to turn ON/OFF frequently
\end{itemize}

\vspace{2mm}
$P_{gmin}$/$Q_{gmax}$ ratio
\begin{itemize}
\item ratio of the generators' two default generation limits: $P_{gmin}$ minimum active power and $Q_{gmax}$ maximum reactive power generation limits
\item each unit receives a value between 0 and 1; after determining the $P_{gmin}$/$Q_{gmax}$ ratio of each generator, individual values are calculated into relative values compared to the maximum of the \textit{time\_period} 
\item smaller ratios are preferred, because those units reduce the reactive power generation imbalance (caused by the significant renewable penetration) more than they contributes toward the active power generation
\end{itemize}

\vspace{1mm}
If the determined GPWD factor of a generator is smaller than 0, it is changed to 0.
Disabled units receive 0 as well.
\vspace{2mm}

\subsection{Universal Selection Scheme} \label{M-One}

GPWD factors are used to rank conventional units, where larger value corresponds to higher rank (greater importance) on the created list.
To decide which units participate in the supply of the demand, the USS method was created.
USS is implemented in each \textit{time\_period} and area separately, and uses the GPWD-ranked list of units.

\vspace{2mm}
Enabled units are selected based on the following values, and after taking the below detailed preparatory steps:
\begin{itemize}
\item Disable every conventional units in the system, then re-enable a unit (the one with the highest $P_{gmax}$ active power generation capability) for each Slack bus.
\item Determine the $hour\_of\_the\_day$ of the \textit{time\_period}.
\item Determine the $renewable\_percentage$ goal value (abbrv.: $renew\_pct$): the planned renewable generation share of total load in the \textit{time\_period} (based on Section \ref{EP}).
\item Calculate active and reactive missing generation of each area: difference between forecasted load and the enabled total renewable generation (based on Section \ref{EP}).
\item Set MW and MVar generation goals in each area:
The area's $MW\_generation\_goal$ is 115\% of the active missing generation (considering the effect of power transmission losses in the system, and keeping 10\% spinning reserve).
The area's $MVar\_generation\_goal$ is the reactive missing generation minus 85\% of the added extra sync-conds' total generation.
\item Take into consideration the enabled units of the Slack buses: deduct (0.5x$P_{gmin}$+0.5x$P_{gmax}$) from their area's $MW\_generation\_goal$, and deduct 50\% of their $Q_{gmax}$ from their area's $MVar\_generation\_goal$. In the equations, $P_{gmin}$, $P_{gmax}$ and $Q_{gmax}$ values are the generation limits of the enabled units.
\end{itemize}

\vspace{2mm}
Generators are enabled until there is missing generation in their area, i.e. the $MW\_generation\_goal$ and/or the $MVar\_generation\_goal$ in their area is greater than 0.

\newpage

General description of the USS method:
\begin{enumerate}
\item take first (or next) generator of the GPWD-ranked list;
\item determine the area and status (can be enabled or must stay disabled) of the unit;
\item decide if the unit needs to be enabled in the \textit{time\_period}; it not, then terminate the setting-process;
\item after enabling the unit, calculate the new area generation goals (deduct the effect of the enabled unit from the old area generation goals);
\item start over from 1).
\end{enumerate}

\vspace{2mm}
The detailed USS method consists of three steps, and further specifies 3) and 4) points of the above general description.
An OPF simulation \cite{opf_1} is performed after each step to determine the success (i.e. power flow convergence) of the created system-setup, and to decide if continuing to the next step is necessary or not.
\vspace{2mm}

\textit{Step 1}: enable as few conventional generators as possible
\begin{itemize}
\item 3): enable units until the $MW\_generation\_goal$ OR $MVar\_generation\_goal$ in their area is greater than 0
\item 4): to get the new generation goals of the enabled unit's area, deduct (0.15x$P_{gmin}$+0.85x$P_{gmax}$) from the old $MW\_generation\_goal$, and deduct (0.85x$Q_{gmax}$) from the old $MVar\_generation\_goal$
\item Once the setting-process ends (either goes through the full GPWD-ranked list, or gets terminated because the generation goal(s) went below 0), the rest of the generators on the list remain disabled in the \textit{time\_period}.
\item If the performed OPF simulation was successful, the status of units is determined and the USS method is concluded; otherwise it proceeds to \textit{Step 2}.
\end{itemize}

\vspace{2mm}
\textit{Step 2}: enable a realistic number of units based on active power generation of conventional generators
\begin{itemize}
\item 3): enable units until the $MW\_generation\_goal$ in their area is greater than 0
\item 4): to get the new area generation goals, this rule applies:
\begin{itemize}
\item \textit when the renewable generation is low, the {time\_period}s require more conventional units (with generation closer to their $P_{gmax}$ limits); when the generation is high, they require less conventional units (with generation closer to their $P_{gmin}$ limits)
\item thus, from the old $MW\_generation\_goal$ deduct:\\
(0.50x$P_{gmin}$+0.50x$P_{gmax}$) if $renew\_pct$$<=$10\%\\
(0.55x$P_{gmin}$+0.45x$P_{gmax}$) if $renew\_pct$$<=$17.5\%\\
(0.60x$P_{gmin}$+0.40x$P_{gmax}$) if $renew\_pct$$<=$25\%\\
(0.65x$P_{gmin}$+0.35x$P_{gmax}$) if $renew\_pct$$>$25\%
\end{itemize}
\item As in \textit{Step 1}, once the setting-process ends, the rest of the generators on the list remain disabled.
\item If the performed OPF simulation was unsuccessful, the method proceeds to \textit{Step 3}.
\end{itemize}

\vspace{2mm}
\textit{Step 3}: enable a realistic number of units based on reactive power generation of conventional generators
\begin{itemize}
\item 3): enable units until the $MVar\_generation\_goal$ in their area is greater than 0
\item 4): to get the new area generation goal, this rule applies:
\begin{itemize}
\item different time of the day requires different number of enabled units: during the night (when reactive power demand is lower) less is needed, while during the day (when reactive power demand is higher) more
\item thus, from the old $MVar\_generation\_goal$ deduct:\\
(0.25x$Q_{gmax}$) if $hour\_of\_the\_day$ = 1-6, 24\\
(0.20x$Q_{gmax}$) if $hour\_of\_the\_day$ = 7-10, 22-23\\
(0.15x$Q_{gmax}$) if $hour\_of\_the\_day $= 11-21
\end{itemize}
\item As in earlier steps, once the setting-process ends, the rest of the generators on the list remain disabled. 
\end{itemize}

\vspace{2mm}
As the USS method is concluded, the operational schedule in the \textit{time\_period} is determined.
Renewable units are enabled and set based on the modifications of Section \ref{EP}.
Conventional units are enabled based on the last performed Step of the USS method, and set to generate with their $P_{gmin}$ and $Q_{gmax}$ values as a starting point. Another performed OPF or PF simulation on the restored power system determines the exact generation setpoints of these units.
\vspace{1mm}

\section{Implemented Algorithms for Result Comparison}


\begin{figure*}[!b]
\centerline{\includegraphics[scale=0.40]{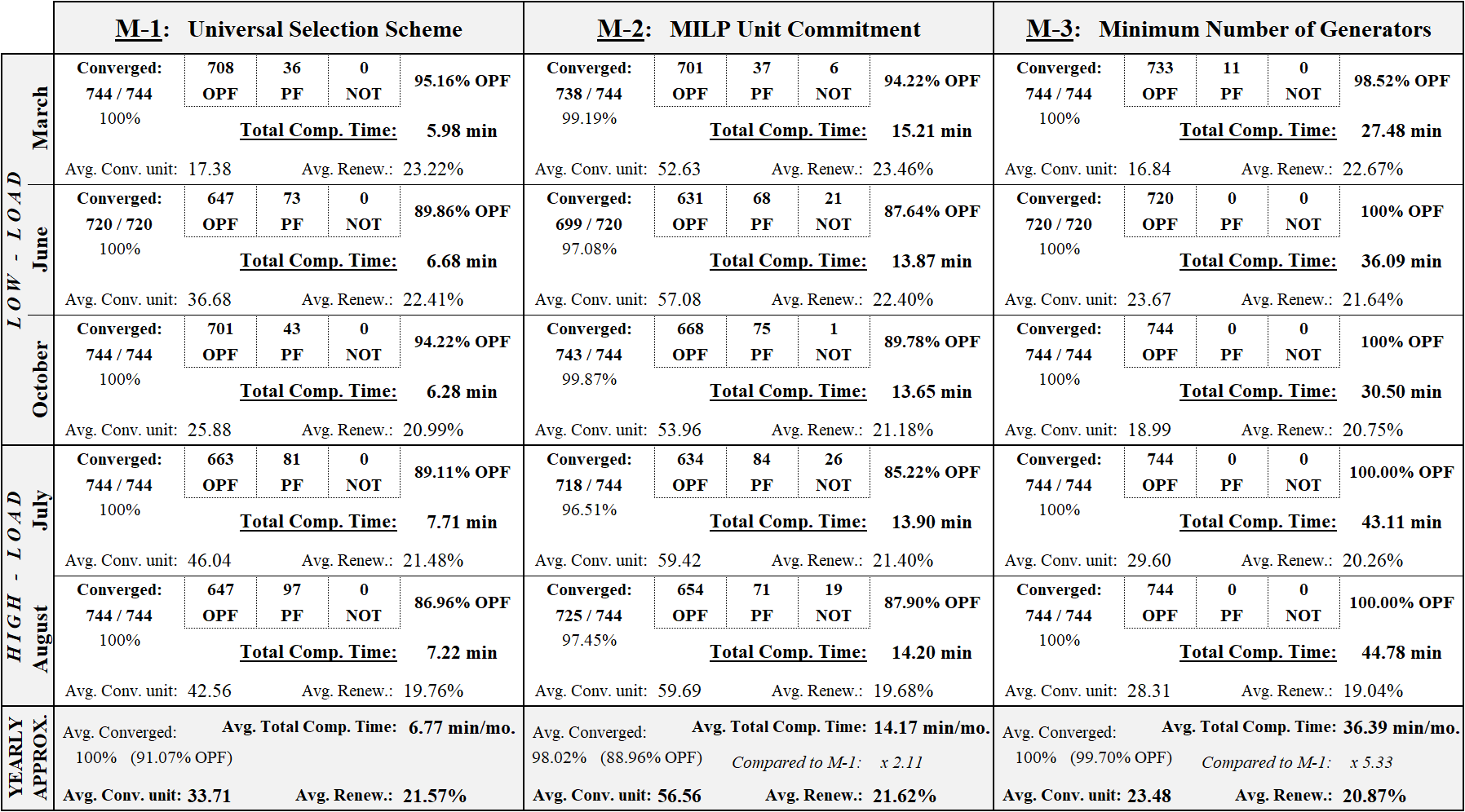}}
\caption{Comparison of different prioritization approaches during normal operation.}
\label{Comparison_normal-operation}
\end{figure*}

\vspace{2mm}

\subsection{MILP Unit Commitment}\label{M-Two}


Unit Commitment is a mathematical optimization problem that determines the optimal operational schedule of generator units within a power system subject to device and operating constraints \cite{uc_1}. In most cases the target objective is to minimize the operational costs throughout the system.

Numerous UC solution approaches have been explored, and algorithms have been developed and tested over the years. Techniques for regulated and deregulated markets, systems with renewable energy resources and energy storage units, distributed generation systems, and more \cite{uc_2} --\cite{uc_3}.
In the electric utility industry, traditionally the Lagrangian Relaxation (LR) technique has been used to solve UC problems, and remains a widely used powerful solution approach \cite{uc_3} --\cite{uc_4}. Nowadays, after the spread of efficient commercial solvers such as CPLEX \cite{uc_5} --\cite{uc_7}, the common and most efficient practice of solving UC problems is through Mixed-Integer Linear Programming (MILP).

MILP algorithms adopt linear programming to solve and check for an integer solution \cite{uc_3},\cite{uc_5}. It is required that the objective function and constraints be a linear function of the decision variables. Their greatest advantage over LR is global optimality; they guarantee a solution that is globally optimal or one with an acceptable tolerance \cite{uc_7} --\cite{uc_9}. On the other hand, they scale poorly and fail when the number of units increases, or when additional modeling detail is integrated. Their efficiency also  suffer from computational delay and the need for large memory \cite{uc_2} --\cite{uc_3},\cite{uc_7}.


\vspace{2mm}
The herein implemented MILP UC algorithm is based upon an openly-accessible UC script by \textit{MathWorks}: \cite{uc_10}. The MILP computation was solved using the INTLINPROG solver of MATLAB's Optimization Toolbox \cite{uc_11}.

\textit{MathWorks}' script was customized and optimized for the used RTS-GMLC system in the following manner:
\begin{itemize}
\item MILP UC was implemented for each system area separately to account for the unique properties of the areas. It is executed in each \textit{time\_period} separately.
\item Only the conventional generators need to be optimized; the data of other units and system elements were ignored.
\item Input data (RTS-GMLC default data) was modified to fit the application circumstances:
\begin{itemize}
\item Fuel cost data was provided in units of [\textdollar/MMBTU].
\item Operational cost data was provided as a piecewise linear cost function with four breaking points. [\textdollar/hr/MW] unit values were calculated for each generator to quicken the algorithm. The given four values were averaged into a single value.
\item When a generator was enabled in the previous \textit{time\_period}, its start-up cost was changed to 0 [\textdollar] in the present \textit{time\_period}.
\item Ramp-up and ramp-down rates were changed from given [MW/min] unit to [MW/hr] units.
\item All values of disabled generators were set to 0, as they are not participating in the algorithm.
\end{itemize}
\item Forecasted load data (targeted MW active power generation of the area) is increased by 5\% to serve as spinning reserve for the generators of the area and to compensate for potential variabilities and modelling inaccuracies.
\item The objective function is the sum of three variables: cost of turning the generator on (Status x Start-up cost), cost of running the generator if it is on ($P_g$ x Operating cost), and cost of generating power ($P_g$ x Fuel cost).
\item The number of integrated modeling details were kept low to increase computational speed.
\end{itemize}
\vspace{2mm}

\subsection{Minimum Number of Generators} \label{M-Three}

The minimum number of conventional generators is the amount of units that is needed to successfully perform an OPF simulation \cite{opf_1} in the created power system, i.e. to reach power flow convergence.
This algorithm determines and enables the minimum number of units in the entire system in each \textit{time\_period} separately.

MNG utilizes the earlier formed GPWD factor-ranked conventional generator list, in which the units are listed from the largest GPWD value unit to the smallest one (presented in Section \ref{GPWD}). In the process -- which is an "enable-and-try" algorithm -- generators are enabled one-by-one, from the top of the list to the bottom, or until the OPF simulation of the resulting system-setup is successful.
\vspace{2mm}

\section{Results and Discussion} \label{Results}

The testing of the proposed prioritization method was done on a computer with an Inter(R) Core(TM) i7-7500U 2.90GHz CPU, and 12GB RAM. The used software was MATLAB R2018b 64-bit, with MATPOWER 6.0 \cite{res_1}.
\vspace{1mm}


\begin{figure*}[!b]
\centerline{\includegraphics[scale=0.175]{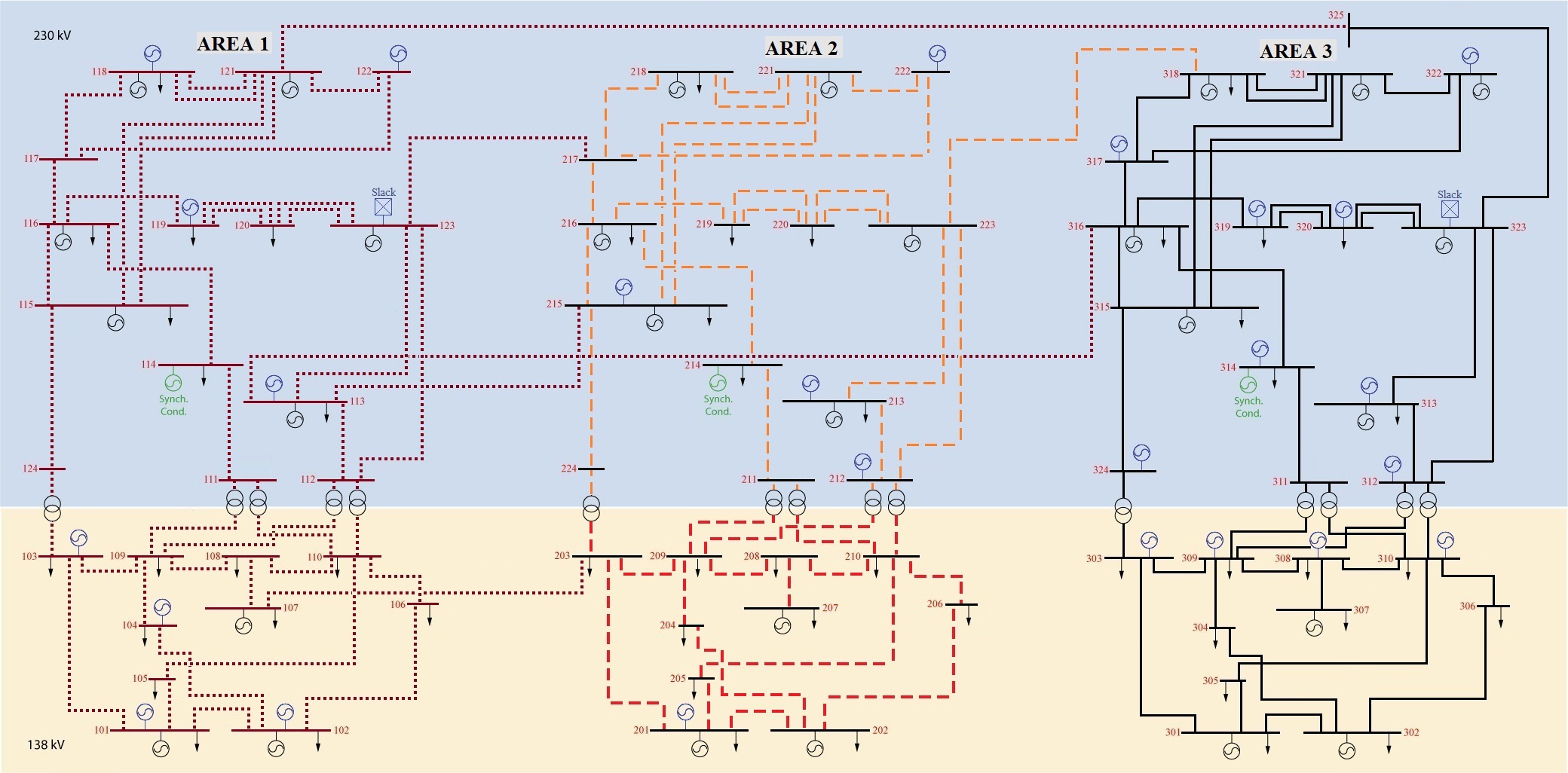}}
\caption{One-line diagram of the RTS-GMLC system during the restoration time-frame.}
\label{RTS-GMLC_CSZ_oneline}
\end{figure*}

\subsection{Prioritization During Normal Operation} \label{Results-Normal}

First, validation experiments were performed during the ``Normal Operation'' of the RTS-GMLC system where all system elements were continuously operational, and connected to the grid according to the above detailed customization changes.
Five months out of year-long data were tested through hourly simulations: the months with the lowest (March, June, and October), and the months with the highest areal and total system loads (July and August).
Each month is 30 or 31 days long, resulting in 720 or 744 simulated \textit{time\_period}s.


Fig.~\ref{Comparison_normal-operation} presents the results in table format. Each column belongs to a different prioritization approach and each row details their performances in a specific month or (in the last row) approximated for the entire year.
The table states the number of \textit{time\_period}s with ``working'' (converged OFP or PF simulation of the created system-setup) and ``not working'' operational schedules, the total computational times, the average number of enabled conventional units, and the average renewable shares of total generation.

As Fig.~\ref{New_Energy_Portfolios} also illustrates, the USS method and the MNG algorithm both determine working operational schedules in every \textit{time\_period}; the MILP UC algorithm, however, is not always able to provide working schedules, which explains the missing (or unrealistic) columns in its graph.
To further validate the conclusion of Section \ref{Observations}, Fig.~\ref{Comparison_normal-operation} proves that the results of all three approaches in yearly average satisfy the 20\% minimum renewable generation requirement.

Comparing the USS method to the MILP UC algorithm, it must be noted that the former dispatches significantly less conventional units, resulting in more feasible and economical schedules.
Although the MNG algorithm creates the best operational schedules, it is the slowest among the three; 5-times slower than the USS method.
Furthermore, even though the implemented MILP UC algorithm was designed to be fast, the proposed prioritization method is 2.11-times faster. Considering the small size of the RTS-GMLC system, this is a significant difference.

\subsection{Prioritization During Restoration} \label{Results-Restoration}

Validation experiments were performed during an ongoing system restoration process. Based on historical data and expected consequences \cite{intro_1},\cite{intro_2}, a fictional restoration time-frame was implemented for a presumed CSZ earthquake event.
To create a connection between the RTS-GMLC system and the PNW, it was assumed that
Area 1 of RTS-GMLC corresponds to the Pacific Coast region of the PNW,
Area 2 to the region between the Coastal Range and the Cascades, and
Area 3 to the region east of the Cascades.


Two weeks data were tested through hourly simulations, between January 26 and February 8. It was assumed that RTS-GMLC operates according to the below schedule (note: ``TP'' is abbreviation of \textit{time\_period}); Fig.~\ref{RTS-GMLC_CSZ_oneline} presents the one-line diagram of the system during this period.
\begin{enumerate}
\item Normal Operation (01/26 TP-1 to 01/26 TP-21)
\item CSZ Earthquake Disaster (01/26 TP-22 to 01/29 TP-9): at 9pm local time a CSZ event struck the region; Area 3 remains intact and continues to operate, while Area 1 and 2 disconnect and enter into complete blackout
\item Partially Restored Operation I. (01/29 TP-10 to 02/03 TP-17): about three days after the CSZ event, the 230 [kV] side of Area 2 (orange dashed lines in Fig.~\ref{RTS-GMLC_CSZ_oneline}) is restored, and is connected to operate with Area 3
\item Partially Restored Operation II. (02/03 TP-18 to 02/08 TP-24): about a week after the CSZ event, the 138 [kV] side of Area 2 (red dashed lines in Fig.~\ref{RTS-GMLC_CSZ_oneline}) is restored and connected to the operating areas; Area 1 (red dotted lines in Fig.~\ref{RTS-GMLC_CSZ_oneline}) remains nonoperational
\end{enumerate}
\vspace{1mm}

\begin{figure*}[htbp]
\centerline{\includegraphics[scale=0.40]{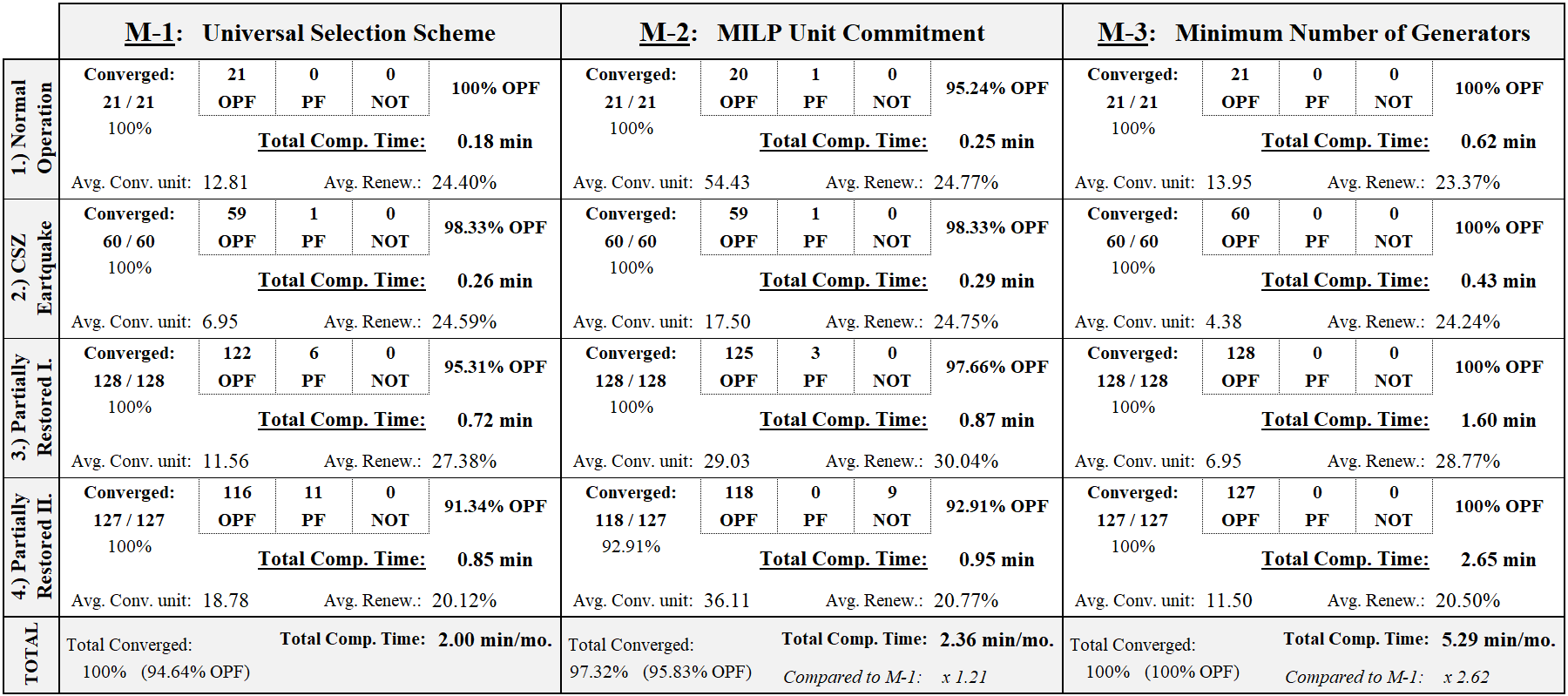}}
\caption{Comparison of different prioritization approaches during the restoration time-frame.}
\label{Comparison_restoration-time-frame}
\end{figure*}

Fig.~\ref{Comparison_restoration-time-frame} presents the results in table format, similarly to Fig.~\ref{Comparison_normal-operation}. The last row displays the total computational times, and the total percentage of ``working'' schedules during the complete restoration time-frame.

The same conclusions can be drawn related to the performances and computational costs as in Section \ref{Results-Normal}.
As was expected, the MILP UC algorithm became much faster as the system size (and element number) was reduced, but the determined schedules are non-feasible in many cases.
The MNG algorithm provided the best schedules during the time-frame, enabled the least amount of conventional units in each step of the restoration, but was considerably slower than other approaches.
The proposed USS method provides the fastest, reliable operational schedules among the three prioritization approaches, regardless if during normal operation or a restoration process.


\subsection{Closing Remarks}

Altogether about six months data were used to perform the validation and benchmarking experiments on the proposed USS method.
The selected periods cover a wide range of possible system-states - months with the highest and lowest areal and system loads, a month from each quarter of the year (to take into account the seasonality of power generation and demand), and times during normal and islanded operation (as part of a restoration process) - all in a power system with significant renewable penetration.

The presented Universal Selection Scheme method (and the associated GPWD factor) proved to be a fast, efficient and convenient tool under various circumstances. Thus, it is advised to be utilized during time-sensitive restoration to rapidly plan the operational schedule of generator units in a power system.

\end{document}